\documentclass[
a4paper,
aps,
prd,
twocolumn,
tightenlines,
preprintnumbers,
nofootinbib,
showkeys,
superscriptaddress
]{revtex4-1}

\usepackage{XCharter}
\usepackage[T1]{fontenc}

\usepackage{fullpage}
\usepackage{amsfonts}
\usepackage{amsmath}
\usepackage{slashed}
\usepackage{amssymb}
\usepackage{graphicx}
\usepackage{epic}
\usepackage{eepic}
\usepackage{epsfig}
\usepackage{latexsym}
\usepackage[dvipsnames]{xcolor}
\usepackage[export]{adjustbox}
\usepackage{float}
\usepackage{multirow}
\usepackage{hyperref}
\usepackage{enumitem}
\hypersetup{colorlinks=true,citecolor=red,linkcolor=NavyBlue,urlcolor=NavyBlue}
\usepackage[caption=false]{subfig}
 
\usepackage{natbib}
\usepackage{tikz} 
\usetikzlibrary{shapes,arrows,positioning,automata,backgrounds,calc,er,patterns}
\usepackage[compat=1.1.0]{tikz-feynman}
\usepackage{relsize}
\usepackage[left=1.7cm,right=1.7cm,top=2.0cm,bottom=2.5cm]{geometry}
\usepackage{shorthand}

\begin{document}
\preprint{IOP/BBSR/2024-06}
\vspace*{0.1cm}
\relscale{0.95}

\title{Boosted top quark inspired leptoquark searches at the muon collider}

\author{Arvind Bhaskar}
\email{arvind.bhaskar@iopb.res.in}
\affiliation{Institute of Physics, Bhubaneswar, Sachivalaya Marg, Sainik School, Bhubaneswar 751005, India}

\author{Manimala Mitra}
\email{manimala@iopb.res.in}
\affiliation{Institute of Physics, Bhubaneswar, Sachivalaya Marg, Sainik School, Bhubaneswar 751005, India}
\affiliation{Homi Bhabha National Institute, Training School Complex, Anushakti Nagar, Mumbai 400094, India}

\date{\today}

\begin{abstract}
The proposed muon collider presents a promising avenue to explore various classes of beyond the Standard Model (BSM) particles. In this paper, we investigate the discovery prospects of the scalar leptoquark (LQ) $S_1$ at a muon collider. We consider two benchmark center-of-mass (C.O.M.) energy scenarios: 5 TeV and 10 TeV. We assume that the LQ decays into a top quark and a muon. The collider analysis for an LQ decaying into a top quark is distinct from that of lighter quarks. A TeV-scale LQ decaying into a top quark can produce an exotic, boosted fat-jet signature. In addition to the usual searches based on pair production of LQs, we also examine the single production mode, which depends on the $S_1 t \mu$ coupling. We demonstrate that systematically combining the pair and single production modes significantly enhances the discovery potential of the LQ at the muon collider. Our signal topology includes at least one hadronically decaying top fat-jet and two oppositely charged muons, thereby enabling the incorporation of the single production mode. We show that even with single production alone, it is possible to probe LQs as heavy as 4.5 TeV (9.0 TeV) in the 5 TeV (10 TeV) C.O.M. scenario for $\mathcal{O}(1)$ couplings.
\end{abstract}

\maketitle


\section{Introduction}
\label{sec:intro}
\noindent
The experimental search for the particle spectrum of standard model (SM) culminated with the discovery of the Higgs boson in 2012 at the large hadron collider (LHC). Despite its tremendous success, the Standard Model (SM) remains an incomplete theory. Although there has been no direct experimental evidence of new physics at colliders, hints of physics beyond the Standard Model (BSM) have emerged in various low-energy experiments. For instance, the measurement of the anomalous magnetic moment of the muon ($(g-2)_{\mu}$) throws a discrepancy of $\approx 4\sigma$ between the SM calculated value and the experimental observation at Fermilab~\cite{Venanzoni:2023mbe, Girotti:2024fdd, Aoyama:2020ynm, Muong-2:2021ojo}. 
Many BSM models have been proposed to explain these low energy experimental anomalies. Leptoquarks (LQs) are one of the promising class of particles to explain such anomalies. LQs are color triplet and electromagnetically charged scalar and vector bosons. They can exist as a weak singlet, doublet or triplet. LQs have been explored extensively in literature ~\cite{Sakaki:2013bfa, Mohanta:2013lsa, Sahoo:2015wya, Dey:2015eaa, Mandal:2015lca, Freytsis:2015qca,Sahoo:2015qha,Sahoo:2015fla,Aydemir:2015oob,Sahoo:2015pzk, Aydemir:2016qqj,Das:2016vkr,Sahoo:2016nvx,
Becirevic:2016oho,Bandyopadhyay:2016oif,Sahoo:2016pet,Faroughy:2016osc,Hiller:2016kry,Bhattacharya:2016mcc,Duraisamy:2016gsd,Das:2017kkm,Assad:2017iib,Dey:2017ede,Biswas:2018jun,Bandyopadhyay:2018syt,Aydemir:2018cbb,Kumar:2018kmr,Angelescu:2018tyl,Mandal:2018kau,Iguro:2018vqb,Aebischer:2018acj,Bar-Shalom:2018ure,Kim:2018oih,Bandyopadhyay:2018syt,Mandal:2018qpg,
Biswas:2018snp,Mandal:2018czf,Biswas:2018iak,Roy:2018nwc,Sahoo:2018ffv,Crivellin:2018yvo,Balaji:2018zna,Fornal:2018dqn,
Alvarez:2018jfb,Mandal:2019gff,Hou:2019wiu,Padhan:2019dcp, Aydemir:2019ynb, Allanach:2019zfr, Zhang:2019jwp, Cornella:2019hct, Baker:2019sli, Bhaskar:2020kdr, Bandyopadhyay:2020klr, Blumlein:1996qp}, but mostly at hadron colliders. LQs are color triplets scalar or vector bosons and carry an electromagnetic charge. They appear in many BSM theories such Pati-Salam models \cite{Pati:1974yy}, SU(5) grand unified theories (GUTs)~\cite{Georgi:1974sy}, models with quark and lepton compositeness~\cite{Schrempp:1984nj}, R-parity violating supersymmetric models~\cite{Barbier:2004ez}. Since, they possess both baryon and lepton numbers, they can combine with quarks and leptons simultaneously. LQs can also play a role in other BSM scenarios such as the production of right handed neutrinos~\cite{Evans:2015ita, Das:2017kkm, Mandal:2018qpg, Bhaskar:2023xkm}, Higgs physics~\cite{Agrawal:1999bk, Bhaskar:2020kdr, FileviezPerez:2021arx, Bhaskar:2022ygp}, dark matter~\cite{Choi:2018stw, Belanger:2021smw}. In our previous works~\cite{Chandak:2019iwj, Bhaskar:2020gkk, Bhaskar:2021gsy}, we investigated the exotic signature of a LQ decaying to a top quark and a lepton at the high luminosity LHC (HL-LHC). 
The ATLAS collaboration has recently published their results of the search for scalar and vector LQs at $\sqrt{13}$ TeV with an integrated luminosity of $139~\rm{fb}^{-1}$ ~\cite{ATLAS:2023prb}. ATLAS has searched for LQs (scalar and vector) decaying to top quarks and light leptons (e, $\mu$). They have excluded a sLQ of mass $1.58~(1.59)$ TeV decaying to a top quark and electron (muon) with $100\%$ branching ratio (BR) with $95\%$ confidence limits (C.L). Similarly, the exclusion limits on its vector counterpart decaying to the same final states are $1.67$ TeV in the minimal coupling scenario and $1.95$ TeV in the Yang-Mills scenario. {The discovery reach for the same final state at the HL-LHC is  only about 1.7 TeV \cite{Chandak:2019iwj}, which motivates to explore the sLQ signal at a lepton collider, where the entire  center of mass (C.O.M)  energy is accessible for resonant production of particles.} In this work, we  study the discovery prospects of scalar LQs (sLQs) decaying to similar final state  at the muon collider~\cite{AlAli:2021let,CERN-LHCC-2015-001}. There are some existing works on the search for LQs at the proposed muon collider~\cite{Ghosh:2023xbj, Bandyopadhyay:2021pld}. A muon collider has certain advantages over a hadron collider. Being a fundamental particle, the muon has the entire C.O.M energy available for collision. It offers high precision measurement of the SM processes. Additionally, the heavier mass of the muon results in reduced energy loss due to synchrotron radiation, allowing for higher energy and luminosity. A muon collider can generate interactions over a range of partonic C.O.M energy $\sqrt{s}$. In this paper, we investigate the discovery reach of the sLQ decaying to a top quark and a muon at the proposed muon collider. We consider the contributions from both the resonant pair and single production of sLQs in our analysis. The single production contribution is included because at higher masses, its contribution falls less rapidly as compared to the pair production process due to less phase space suppression. We also discuss about an additional nonresonant mode of LQ production leading to the same final states. We systematically incorporate all the relevant production modes of the sLQ that contribute to the desired final state involving a pair of top quarks and dimuons. We consider the hadronic decay of the top quark which forms an exotic boosted fatjet signature. To perform the analysis, we consider two benchmark C.O.M energies-- $5$ TeV and $10$ TeV and their corresponding integrated luminosities at $3~\rm{ab^{-1}}$ and $10~\rm{ab^{-1}}$ respectively. We find that a muon collider with the above mentioned C.O.M energy outperforms the discovery reach of a sLQ decaying to a similar final state at the HL-LHC. We highlight its capabilities in probing higher energy scales and achieving greater precision in the study of new physics.

The paper is organised as follows: we introduce the sLQ model in section~\ref{sec:model}. We discuss the decay widths and branching ratios  in section~\ref{sec:decay}, and explain the search strategy in section~\ref{sec:pheno}. Finally, we present our results and conclude in section~\ref{sec:result}.


\section{Theoretical Set Up}
\label{sec:model}
In this paper, we consider a weakly singlet scalar leptoquark $S_1=(\overline{\mathbf{3}},\mathbf{1},1/3)$. Following the notation of Ref.~\cite{Dorsner:2016wpm}, the interaction terms of the $S_1$ Lagrangian can be written as follows:


\begin{align}
\label{eq:LagS1}
\mathcal{L} \supset &~ y^{LL}_{1\,ij}\bar{Q}_{L}^{C\,i} S_{1} i\sigma^2 L_{L}^{j}+y^{RR}_{1\,ij}\bar{u}_{R}^{C\,i} S_{1} \ell_{R}^{j}+\textrm{H.c.},
\end{align}
where $u_{R}$ and $\ell_{R}$ are a SM right-handed up-type quark and a charged lepton, respectively. $Q_L$ and $L_L$ are the SM left-handed quark and lepton doublets, respectively. The superscript $C$  denotes  
charge conjugation and $\sigma^2$ is the second Pauli matrix. The generation indices are denoted 
by $i,j=\{1,\ 2,\ 3\}$. The color indices are suppressed. We write the neutrinos collectively as $\nu$ since the
LHC is neutrino flavours-blind. We focus on $S_1$, that exclusively interacts with a second generation lepton and a third generation quark. Hence, the above Lagrangian simplifies to the following terms,
\begin{align}
\mathcal{L} \supset &\ y^{LL}_{1\ 32} \left(-\bar{b}_{L}^C \nu_{\mu}+\bar{t}_{L}^{C}\mu^j_{L} \right)S_{1}
+y^{RR}_{1\,32}\ \bar{t}_{R}^{C} \mu^j_{R} S_{1}+\textrm{H.c.}
\label{eq:LagS1us}
\end{align}
Note that, in the above the superscript $LL/RR$ on the coupling $y_1$ represents the chirality of the $\mu$ and $t$. In the rest of the work, we shall use $y_1$ as a generalized notation for our new coupling and specify the chirality when required.


\begin{figure*}[htb]
\captionsetup[subfigure]{labelformat=empty}
\subfloat[\quad\quad\quad(a)]{\includegraphics[width=0.25\linewidth]{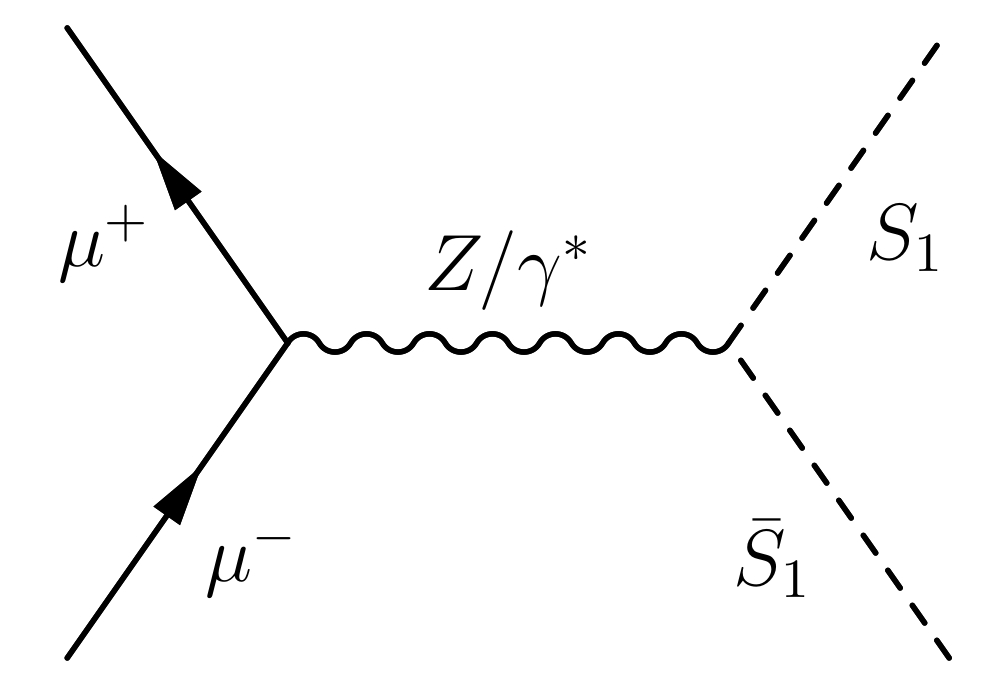}\label{fig:pairproduction2}}
\subfloat[\quad\quad\quad(b)]{\includegraphics[width=0.25\linewidth]{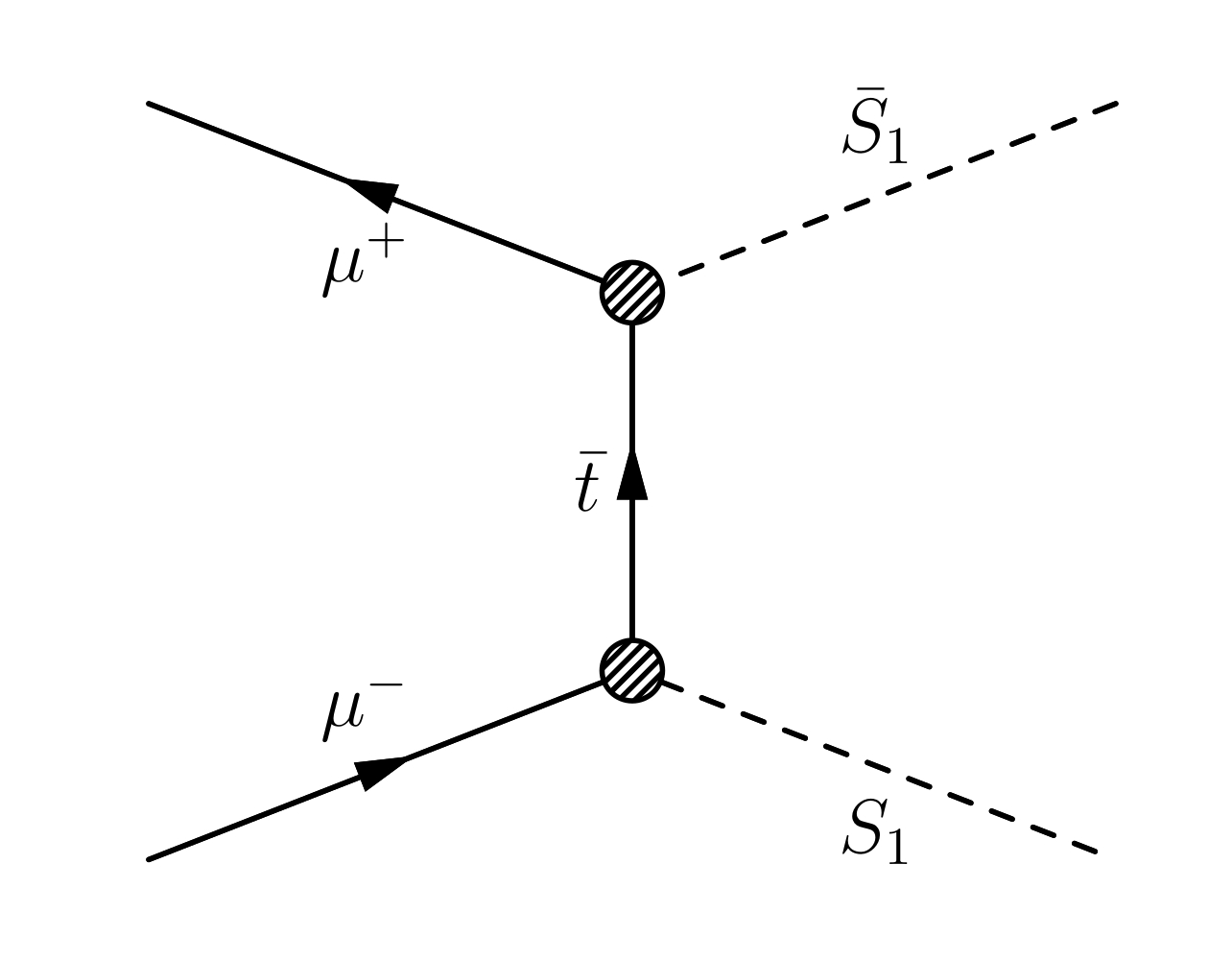}\label{fig:pairProduction}}
\subfloat[\quad\quad\quad(c)]{\includegraphics[width=0.25\linewidth]{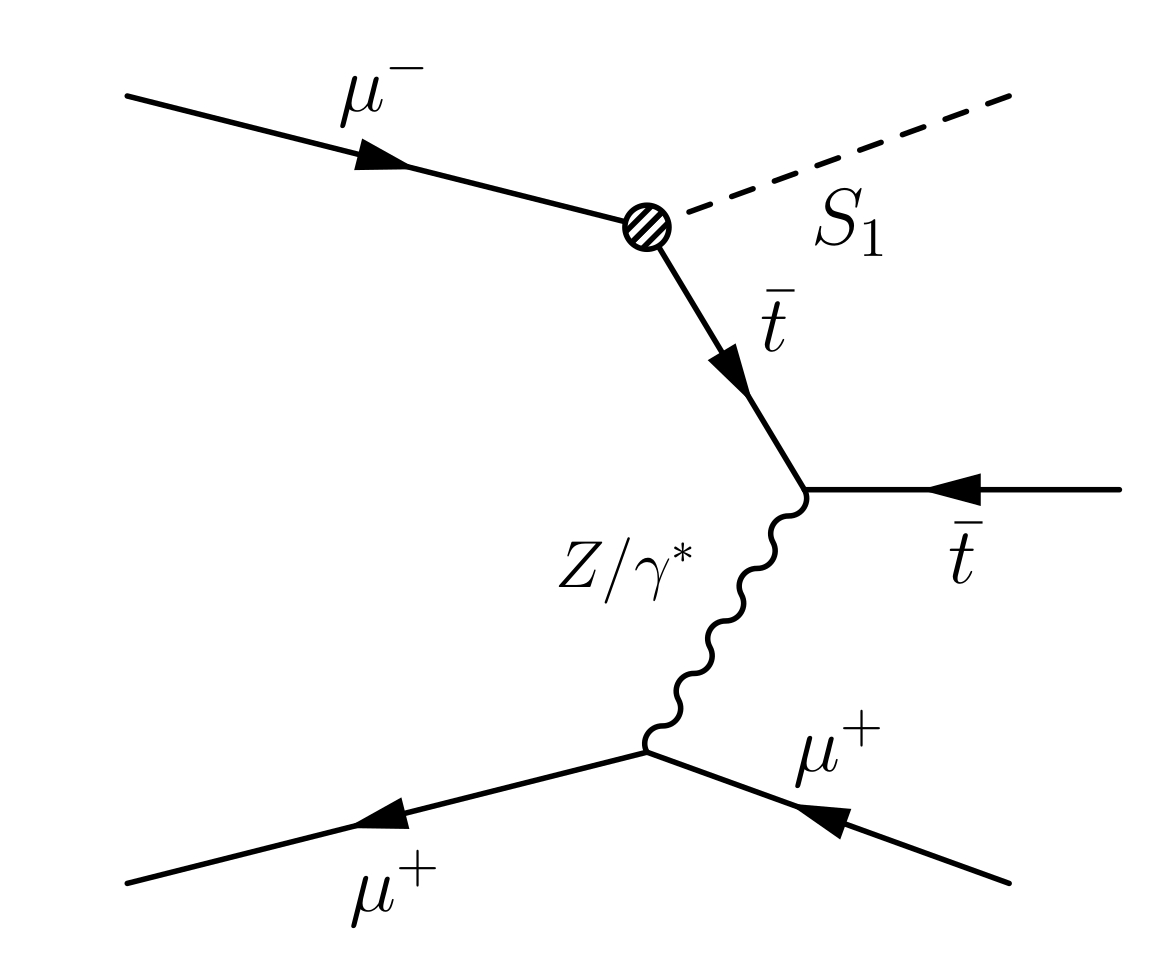}\label{fig:singleproduction}}
\subfloat[\quad\quad\quad(d)]{\includegraphics[width=0.25\linewidth]{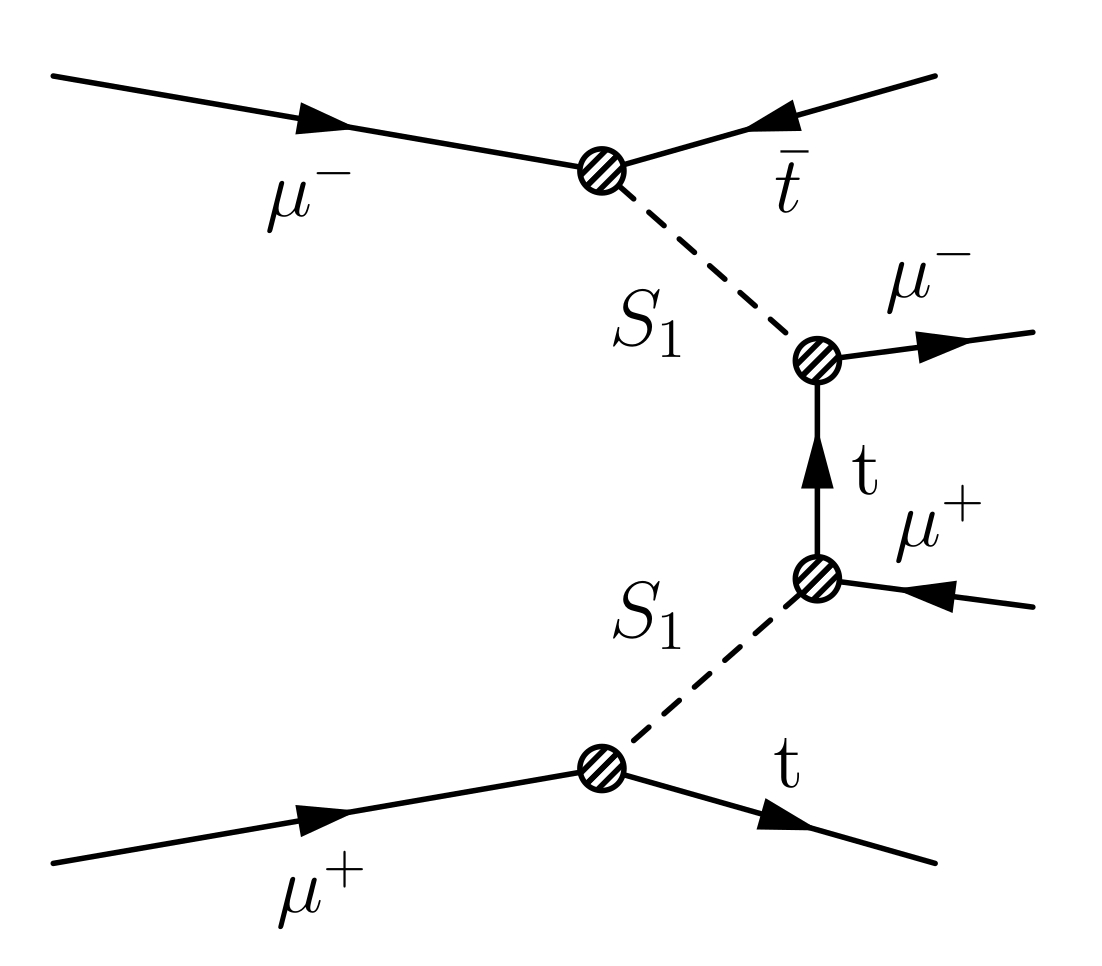}\label{fig:Newphysics}}\\
\caption{We show the representative Feynman diagrams for the $S_1$ production at the muon collider. (a) and (b) correspond to the pair production mode. (c) shows the single production mode. (d) stands for the nonresonant production mode.}
\label{fig:feynman}
\end{figure*}

\section{Decay width and Branching Ratios}
\label{sec:decay}
\noindent
If we consider only a nonzero $y^{LL}_{1 32}$  in Eq.~\eqref{eq:LagS1us}, then  $S_1$ can decay to a top quark and a muon and it can also decay to b quark and a neutrino with a BR $\approx~50\%$. Whereas, if one considers only the  $y^{RR}_{1 32}$ to be nonzero then the $S_1$ decays to a top quark and a muon with $100\%$ BR. We show the  expressions of the decay widths for $S_1$ sLQ as a function of its mass $M_{S_1}$ and new coupling $y_1$.
\begin{equation}
    \Gamma(S_1 \to t \mu)=  \frac{(M_{S_1}^2 - m_t^2)^2 [(y^{LL}_{1\ 32})^2 + (y^{RR}_{1\ 32})^2]}{16\pi M_{S_1}^2} \end{equation}  
    \begin{equation}
\Gamma (S_1 \to b \nu)= \frac{(M_{S_1}^2 - m_b^2)^2 (y^{LL}_{1\ 32})^2 }{16\pi M_{S_1}^2}
\label{eq:partialwidths}
    \end{equation}


\begin{figure*}[]
\captionsetup[subfigure]{labelformat=empty}
\subfloat[\quad\quad\quad(a)]{\includegraphics[width=0.35\linewidth]{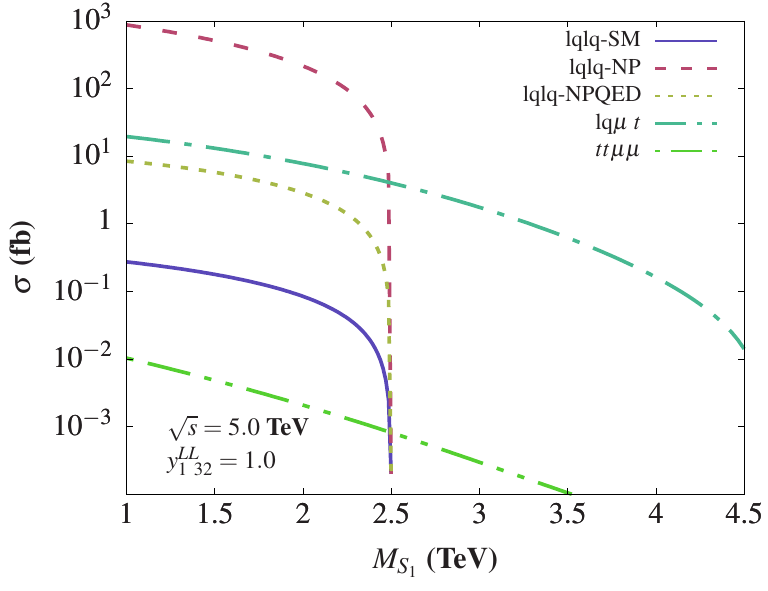}\label{fig:MS13_CS_LH_05}}
\subfloat[\quad\quad\quad(b)]{\includegraphics[width=0.35\linewidth]{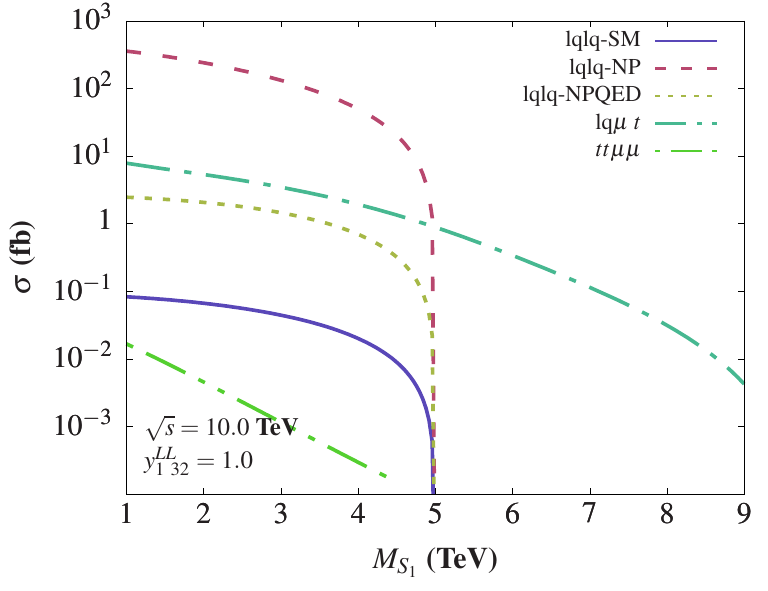}\label{fig:MS13_CS_LH_10}}\\
\subfloat[\quad\quad\quad(c)]{\includegraphics[width=0.35\linewidth]{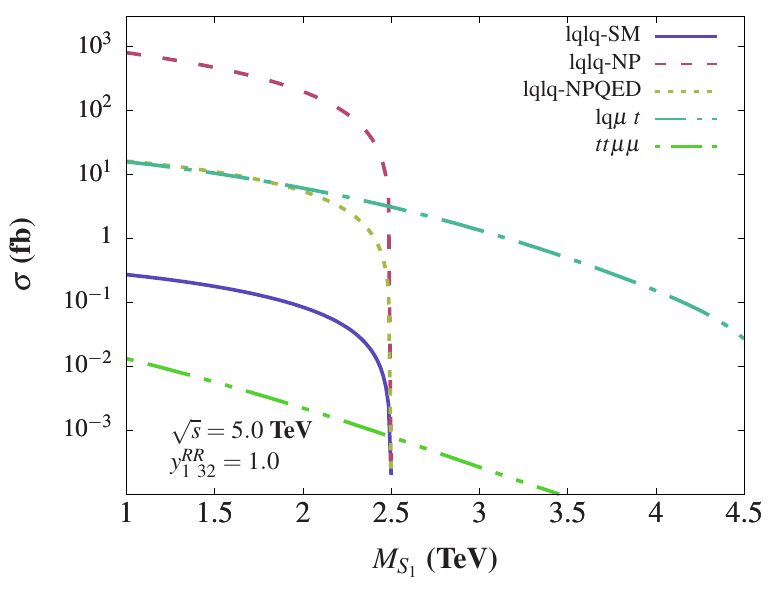}\label{fig:MS13_CS_RH_05}}
\subfloat[\quad\quad\quad(d)]{\includegraphics[width=0.35\linewidth]{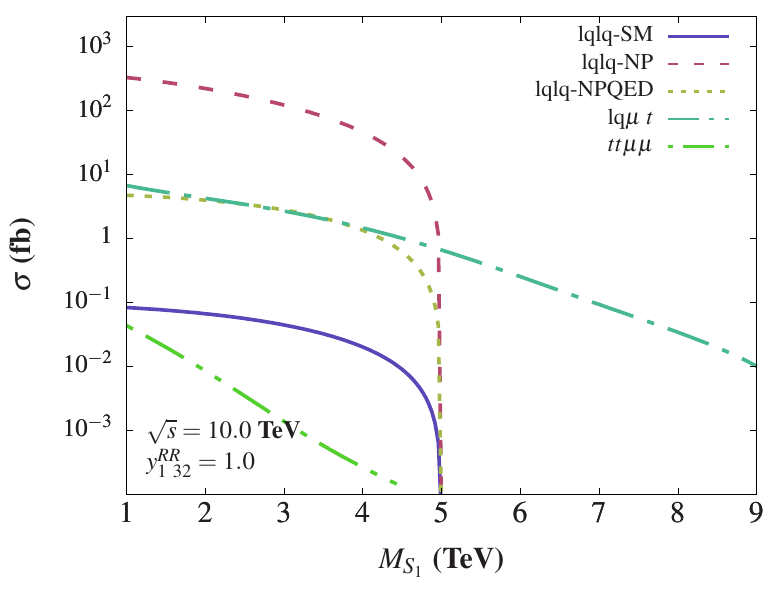}\label{fig:MS13_CS_RH_10}}
\caption{The production  cross section of $S_1$ for  illustrative values of $y^{LL}_{1~32}$ and $y^{RR}_{1~32}$ at $\sqrt{s}=5.0$ TeV [(a) and (c)] and for $y^{LL}_{1~32}$ and $y^{RR}_{1~32}$ at $\sqrt{s}=10.0$ TeV [(b) and (d)]. We have presented the contributions from both the resonant pair and single production modes, as well as the non-resonant production mode. For all processes dependent on the Yukawa coupling $y_1$, we have adopted the benchmark value of $y_1 = 1.0$.}
\label{fig:SLQCS}
\end{figure*}


\section{Search strategy of sLQ at the muon collider}
\label{sec:pheno}
We use the {\tt FeynRules}~\cite{Alloul:2013bka} software package to encode the Lagrangian mentioned in Eqs.~\eqref{eq:LagS1us} to generate the universal Feynman output (UFO) model file. We generate the signal and background events using {\tt MadGraph5aMC@NLO-v3.5.3}~\cite{Alwall:2014hca} at the leading order (LO). The generated signal and background events are passed through {\tt Pythia8}~\cite{Sjostrand:2006za} for showering and hadronization. For simulating the detector effects we use {\tt Delphes3}~\cite{deFavereau:2013fsa} with Delphes ILD card. We cluster the fatjets coming from the decay of top quarks using Cambridge-Achen~\cite{Dokshitzer:1997in} clustering algorithm (with $R = 1.5$) in {\tt FastJet}~\cite{Cacciari:2011ma}.

\subsection{Production at the Muon collider}
The LQs can be produced resonantly as a pair or singly along with a top quark and a muon. In Figure.~\ref{fig:pairproduction2}, we show the representative Feynman diagram for pair production of $S_1$,  mediated by a $Z$ boson or a photon. This contribution doesn't depend on the new coupling $y_1$. Figure.~\ref{fig:pairProduction} depicts the $S_1$ pair production via a $t$-channel top quark exchange. The cross-section contribution from this $t$-channel diagram is proportional to the fourth power of the new coupling constant--$y_1^4$. Additionally, there is an destructive interference between these two diagrams. The cross section from this interference term scales as $y_1^2$. Once produced, the sLQ decays to either a top quark and a charged lepton or a bottom quark and a neutrino depending on the chirality of the yukawa coupling $y_1$, { leading to the following final states.}
\begin{eqnarray}
\mu^-\mu^+\ \to\left\{\begin{array}{ccl}
S_1\bar{S_1}&\to& (t\mu) (t\mu)/(t \mu)(b\n)\ /\ (b\n)(b\n)\\
\end{array}\right\}.
\label{eq:pair_all}
\end{eqnarray}
For the  single production of LQs (see Fig.~\ref{fig:singleproduction}), the cross section contribution scales as $y_1^2$. For single production, the subsequent decay of $S_1$ leads to the following final states, 
\begin{eqnarray}
\mu^-\mu^+&\ \to&\left\{\begin{array}{ccl}
S_1 t \mu &\to & (t\mu)t\mu \\
S_1 t \mu &\to & (b\nu)t \mu \\
S_1 b \nu &\to & (t \mu) b\nu \\
S_1 b \nu &\to & (b\nu) b\nu 
\end{array}\right\},\label{eq:single1}\\
\end{eqnarray}
We also consider the processes such $\mu^+ \mu^- \to S_1 S^*_1 \to S_1 t \mu$, that could contribute to the single production mode. In addition to these two resonant $S_1$ production modes leading to ditop and dimuon final state, there is an additional nonresonant production mode that leads to the same final state. { In Fig.~\ref{fig:Newphysics}, we show the respective  Feynman diagram for this process.} { The cross section for this $t$-channel exchange contribution scales as $y^8_1$.} For our discovery prospect studies, we ignore the contribution from this mode as the cross section is small compared to the pair and single production modes. But this mode can become important for higher LQ masses and large $y_1$ couplings. { In this paper, we focus on the following final state signature:}
\begin{eqnarray}
\mu^-\mu^+ \to S_1 \bar{S_1}, S_1 t \mu \to (t\mu) (t\mu)
\label{finalsignal}
\end{eqnarray}
In Fig.~\ref{fig:SLQCS}, we have shown the cross section for the $S_1$ resonant (pair and single) and non-resonant production modes as a function of its mass for the choice  of new couplings $y_1=1.0$ and C.O.M energies ($\sqrt{s}=5, 10$ TeV). The legend mentioned in Fig.~\ref{fig:SLQCS} is explained as follows. \emph{lqlq-SM} corresponds to the pair production contribution from the Feynman diagram in Fig.~\ref{fig:pairproduction2}. \emph{lqlq-NP} corresponds to the contribution from Fig.~\ref{fig:pairProduction}. \emph{lqlq-NPQED} corresponds to the contribution coming from the interference between the diagrams in Fig.~\ref{fig:pairproduction2} and Fig.~\ref{fig:pairProduction}. $lq\mu t$ stands for the contribution from the single production mode (Fig.~\ref{fig:singleproduction} and other single production diagrams such as, $\mu^+ \mu^- \to \mu^+ \mu^{-*} \to S_1 t \mu$) and $tt\mu\mu$ shows the cross section contribution from the nonresonant mode (Fig.~\ref{fig:Newphysics}).
In Figs~\ref{fig:MS13_CS_LH_05} and \ref{fig:MS13_CS_RH_05}, we show the production cross section of sLQs for couplings $y^{LL}_{1~ 32}$ and $y^{RR}_{1~ 32}$, respectively at C.O.M $\sqrt{s}=5$ TeV. Similarly in Figs.~\ref{fig:MS13_CS_LH_10} and ~\ref{fig:MS13_CS_RH_10}, we show the cross section for left handed and right handed Yukawa coupling at C.O.M $\sqrt{s}=10$ TeV. We consider the benchmark coupling $y^{LL/RR}_{1~ 32} = 1.0$ for all of the  scenarios. { As stated above, once produced, we primarily consider the decay of sLQ to top quark and muon and ignore the final states containing a $b$ quark and neutrino. The top quark produced from a TeV scale sLQ will form a boosted fatjet and shall be accompanied by a high $p_T$ muon. In the next sub-section, in defining our analysis strategy, we keep this into consideration, and therefore, we demand the presence of at least one boosted top quark forming a fatjet in the final state.}


\subsection{Signal selection and backgrounds processes}
We consider at least one hadronically decaying top quark that forms a fatjet and exactly two high-$p_{\rm T}$ opposite sign muons as our signal. The motivation for choosing such a signal is to include the contributions from the single production mode as well. We consider the hadronic decay of the top quark, as the decay of the $W$ boson (originating from the top quark) to jets is more probable than its leptonic decay. Moreover, being  a muon collider, the hadronic final state containing a high $p_T> 500 $ GeV fat-jet may over-rule similar final state generated from relatively smaller SM background.
We list the appropriate SM processes that can contribute as background processes in Table~\ref{tab:Backgrounds}, and present a brief discussion below.  
\begin{enumerate}
    \item $V +$ jets : $V= W, Z$. The SM $W$ and $Z$ boson production plus additional jets can act as potential background processes that can lead to the desired final state.

    \begin{itemize}
        \item[--] $W$ + jets: We generate this background by simulating the process $\mu^- \mu^+ \rightarrow W^{\pm} +$ jets. The $W$ decays to a muon and a neutrino. The additional jets can mimic the top-like fatjet. The second muon can come from a jet misidentified as a lepton.
        \item[--] $Z$ + jets: The $Z$ background is obtained similarly. The $Z$ boson decays leptonically into a dimuon final state. The top-like fatjet is obtained from combining the additional jets.

    \end{itemize}
    
    \item $t\bar{t} +$ jets : The ditop production is one of the dominant background processes. Here, we decay both the top quarks to muons and neutrinos. The top-like fatjet can be obtained by the additional QCD jets and the $b$ quarks. 

    \item $VV + $ jets: There are two types of SM diboson processes which can serve as background. 
        \begin{itemize}
            \item [--] $W~W$ + jets: Here, we generate two $W$ bosons and decay both of them to dimuons and neutrinos in the final state. The top-like fatjet can be obtained by combining the additional QCD jets.
            \item[--] $Z_{\ell}Z_h$ + jets: Here, one of the $Z$ decays to dimuons and the other one decays hadronically to QCD jets. Combining the hadronic decays of the $Z$ boson with the additional QCD jets, one could reconstruct the top-like fatjet.
        \end{itemize}
    \item $t\bar{t} \mu^- \mu^+$ : We decay the top quarks hadronically to $b$ quarks and $W$ bosons. The $W$ boson further decays to a pair of QCD jets.
\end{enumerate}

For better statistics, we apply some generational level cuts on our background processes. We list the generational level cuts.
\begin{enumerate}
\item $p_{\rm T}(\mu_1)> 80$ GeV,
\item $p_{\rm T}(j_1)> 100$ GeV,
\item Invariant mass $M(\mu_1,\mu_2) > 110$ GeV ($Z$-mass veto),
\item $\rm{H}_T > 500$ GeV.
\end{enumerate}

Here $\rm{j}_i$ and $\mu_i$ denotes the $i^{\rm th}$ $p_{\rm T}$-ordered jet and muon respectively. After generating events with the above generation-level cuts, we apply the following final selection criteria sequentially on the signal and the background events.


\begin{table}[t!]
\begin{center}
\begin{tabular}{|c|c|c|c|}
\hline
\multicolumn{2}{|c|}{Background } & $\sg (10.0)$ TeV  & $\sg (5.0)$ TeV\\ 
\multicolumn{2}{|c|}{processes}&(pb)&(pb)\\\hline\hline
$V + jets$  & $Z + jets$  &  $1.34 \times 10^{-7}$  &  $1.04\times 10^{-6}$\\ \cline{2-4} 
                  & $W + jets$  & $0.0001915$  & $0.001063$\\ \hline
$VV + jets$  & $WW + jets$  & $0.0001666$  & $5.2\times 10^{-5}$\\ \cline{1-4} 
$tt+jets$ & --  & $9.9 \times 10^{-5}$  & $6.5 \times 10^{-5}$\\ \cline{1-4}
$tt\mu\mu$ & --  &  $3.65 \times 10^{-7}$  &  $8.0 \times 10^{-7}$\\ \hline
\end{tabular}
\caption{We list the dominant background processes for our analysis. These cross sections have been obtained for C.O.M energies $\sqrt{s}=10$ TeV and $\sqrt{s}=5$ TeV.}
\label{tab:Backgrounds}
\end{center}
\end{table}

\subsection{Discovery potential}
\label{sec:dispot}
Below, we explain the cut flow algorithm for signal selection and the list of cuts applied on the signal and background events.
We mention the cuts for the benchmark scenario $\sqrt{s}=10$ TeV and coupling $y^{LL}_{1~ 32} = 1.0$. Similar steps and cuts have been followed for the other benchmarks scenarios. 
\noindent
\begin{itemize}
    \item [--] Two opposite sign--muons, with atleast one muon with a high ${p}_T(\mu_1) > 300$ GeV and pseudorapidity $|\eta(\mu)|<2.5$.
    \item [--] Invariant mass of the lepton pair $M(\mu_1,\mu_2) > 200$ GeV.

    \item [--] Atleast one $b$ quark with a ${p}_T(b) > 200$ GeV.

    \item[--] Atleast one top-like fatjet with mass of the fatjet, $160 < \rm{m}_{fj} < 200$ GeV. Here, fj implies fatjet.

    \item[--] Transverse momentum of the top-like fatjet ${p}^{fj}_T > 500$ GeV. Scalar sum of transverse momentum $\rm{H}_T > 900$ GeV.
\end{itemize}

{\renewcommand\theenumi{\bfseries $\mc C_\arabic{enumi}$:}
\renewcommand\labelenumi{\theenumi}

We calculate the statistical significance $\mc{Z}$~\cite{Cowan:2010js} using the following formula,
\begin{align}
\mc{Z} = \sqrt{2\lt(N_S+N_B\rt)\ln\lt(\frac{N_S+N_B}{N_B}\rt)-2N_S}\, ,\label{eq:sig}
\end{align}
where $N_S$ and $N_B$ are the numbers of the signal and background events, respectively, surviving the selection cuts as mentioned above.
\begin{gather}
\label{eq:Numevents}
N_S = (\sg_{\rm pair} \times \epsilon_{\rm pair} \times \beta^2 + y^2_1 \sg_{\rm single} \times \epsilon_{\rm single} \times \beta) \times \mathcal{L}\\
N_B = \sg_{\rm B} \times \epsilon_{B} \times \mathcal{L},    
\end{gather}

Here, we define $\sigma_{pair}$ as,
\begin{equation}
\sigma_{pair} = \sigma_{SM} + y^2_1 \sigma_{NPQED} + y^4_1 \sigma_{NP}
\label{eq:paircontribution}
\end{equation}

\noindent
where $\sigma_{\rm pair}$ is the cross section from the combined pair production mode. $\sigma_{SM}$, $\sigma_{NP}$ and $\sigma_{NPQED}$ are the cross sections from Feynman diagrams in the Figs.~\ref{fig:pairproduction2}, \ref{fig:pairProduction} and the interference of the two respectively. $\sigma_{\rm single}$ is the cross section from the single production process, $\sigma_{\rm B}$ is the cross section of the background process, $\epsilon_{x}$ denotes the fraction of events surviving the cuts mentioned above, $\beta$ is the appropriate BR and $\mathcal{L}$ is the luminosity of the muon collider. $\epsilon_{pair}$ remains similar for the different pair production contributions mentioned in Eq.~\eqref{eq:paircontribution} . For the benchmark scenario $\sqrt{s}=10$ TeV, we have assumed the luminosity to be $10~\rm {ab}^{-1}$ and for $\sqrt{s}=5$ TeV, we have assumed the luminosity to be $3~\rm {ab}^{-1}$.


\begin{figure*}[]
\captionsetup[subfigure]{labelformat=empty}
\subfloat[\quad\quad\quad(a)]{\includegraphics[width=0.35\linewidth]{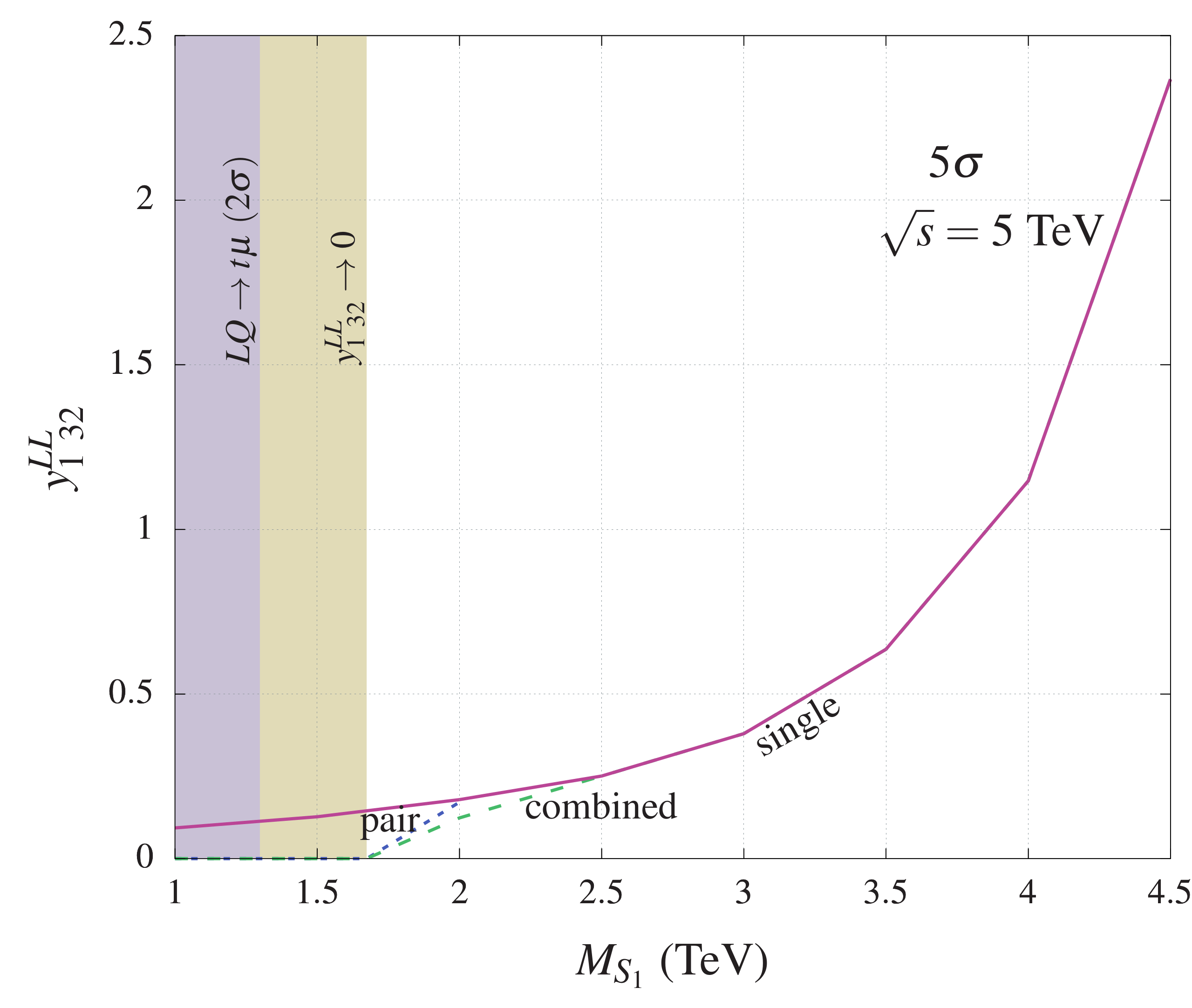}
\label{fig:MS1LQ_LM32L_single_5sig_5TV}}
\subfloat[\quad\quad\quad(b)]{\includegraphics[width=0.35\linewidth]{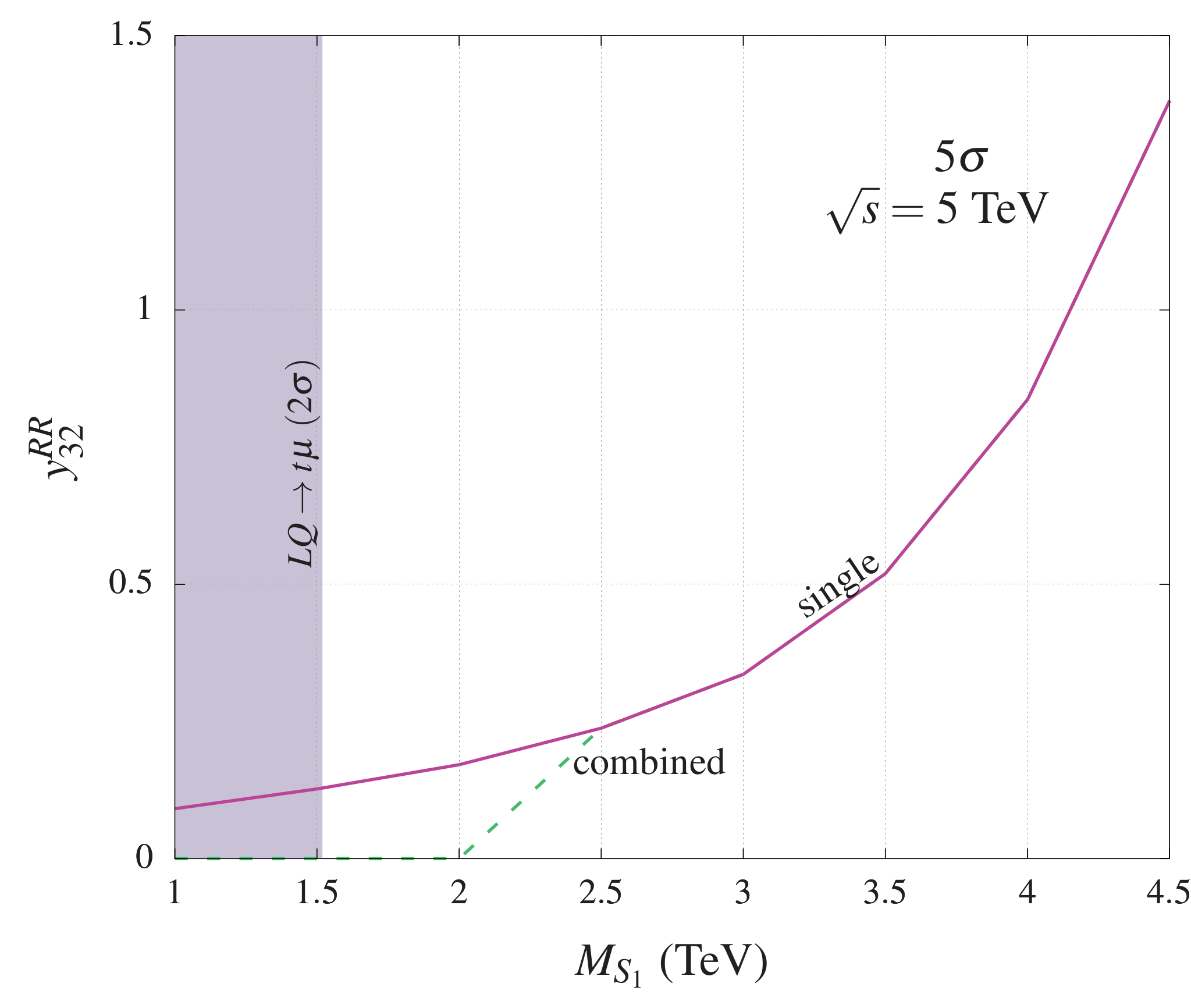}
\label{fig:MS1LQ_LM32R_single_5sig_5TV}}\\
\subfloat[\quad\quad\quad(c)]{\includegraphics[width=0.35\linewidth]{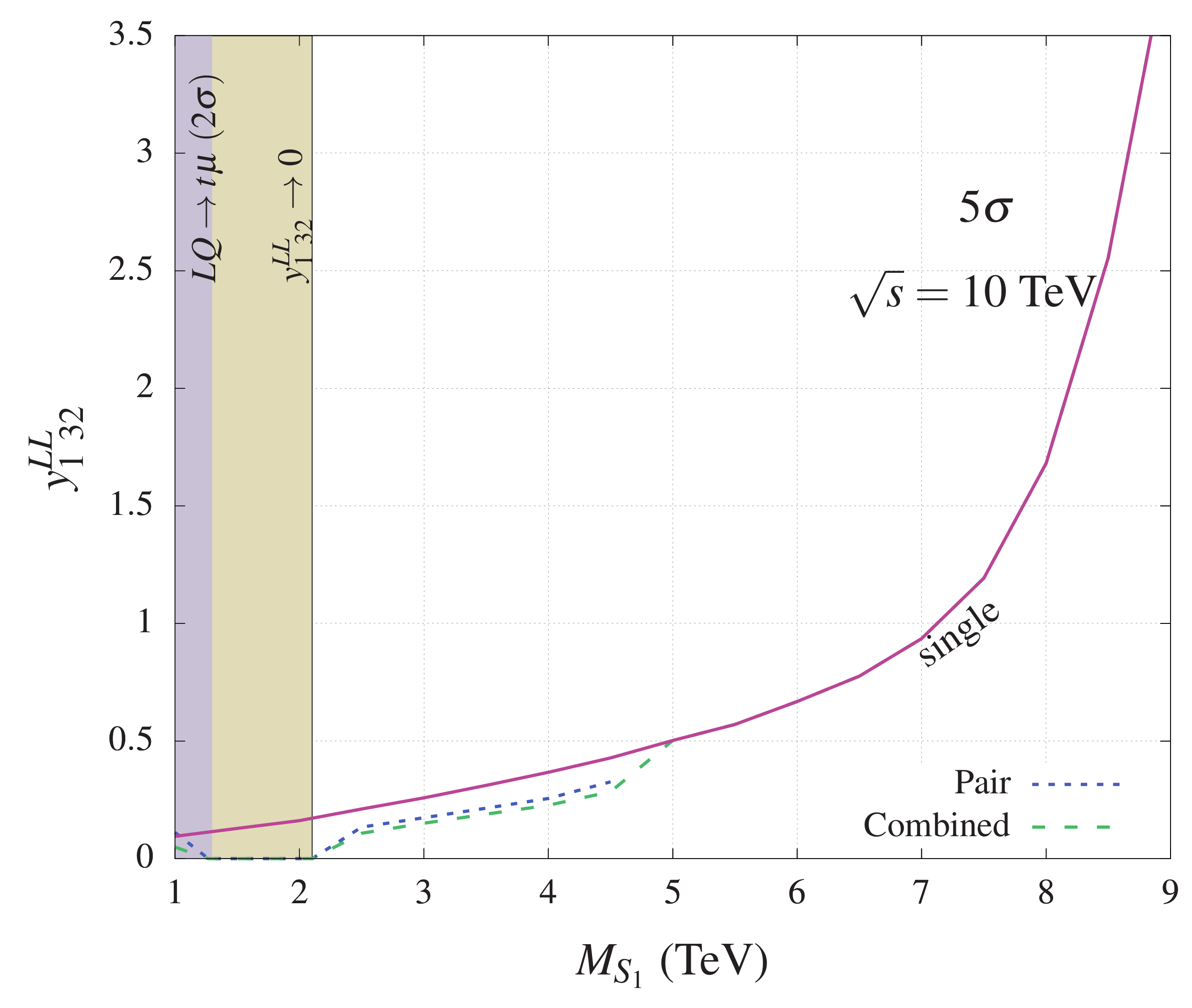}\label{fig:MS1LQ_lambda_32L_5sig_10TeV}}
\subfloat[\quad\quad\quad(d)]{\includegraphics[width=0.35\linewidth]{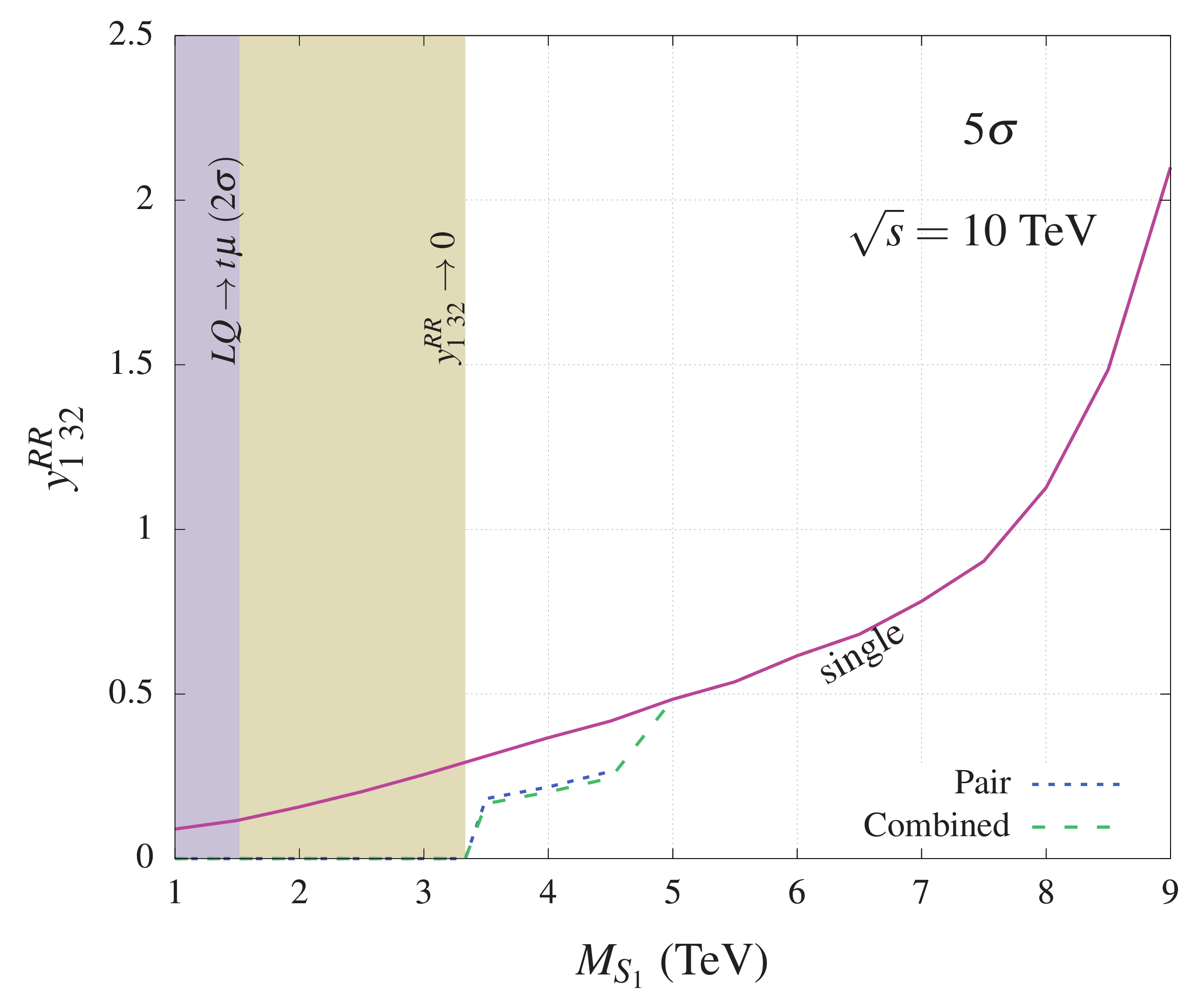}\label{fig:MS1LQ_lambda_32R_5sig_10TeV}}
\caption{The $5\sigma$ discovery reach contours as a function of the new yukawa coupling $y_1$ and $M_{S_1}$ (TeV) for benchmark C.O.M energies $5.0$ TeV [(a) and (b)] and $10.0$ TeV [(c) and (d)]. These plots show the smallest coupling $y_1$ required to obtain a $5\sigma$ discovery reach for a range of $S_1$ mass at an integrated luminosity of $3\rm{ab^{-1}}$ and $10\rm{ab^{-1}}$ respectively. The yellow color shows the discovery reach if one assumes the coupling independent pair production contribution only. The violet color shows the region excluded by the LHC searches with $95\%$ C.L.}
\label{fig:5sigma}
\end{figure*}


\begin{figure*}[]
\captionsetup[subfigure]{labelformat=empty}
\subfloat[\quad\quad\quad(a)]{\includegraphics[width=0.35\linewidth]{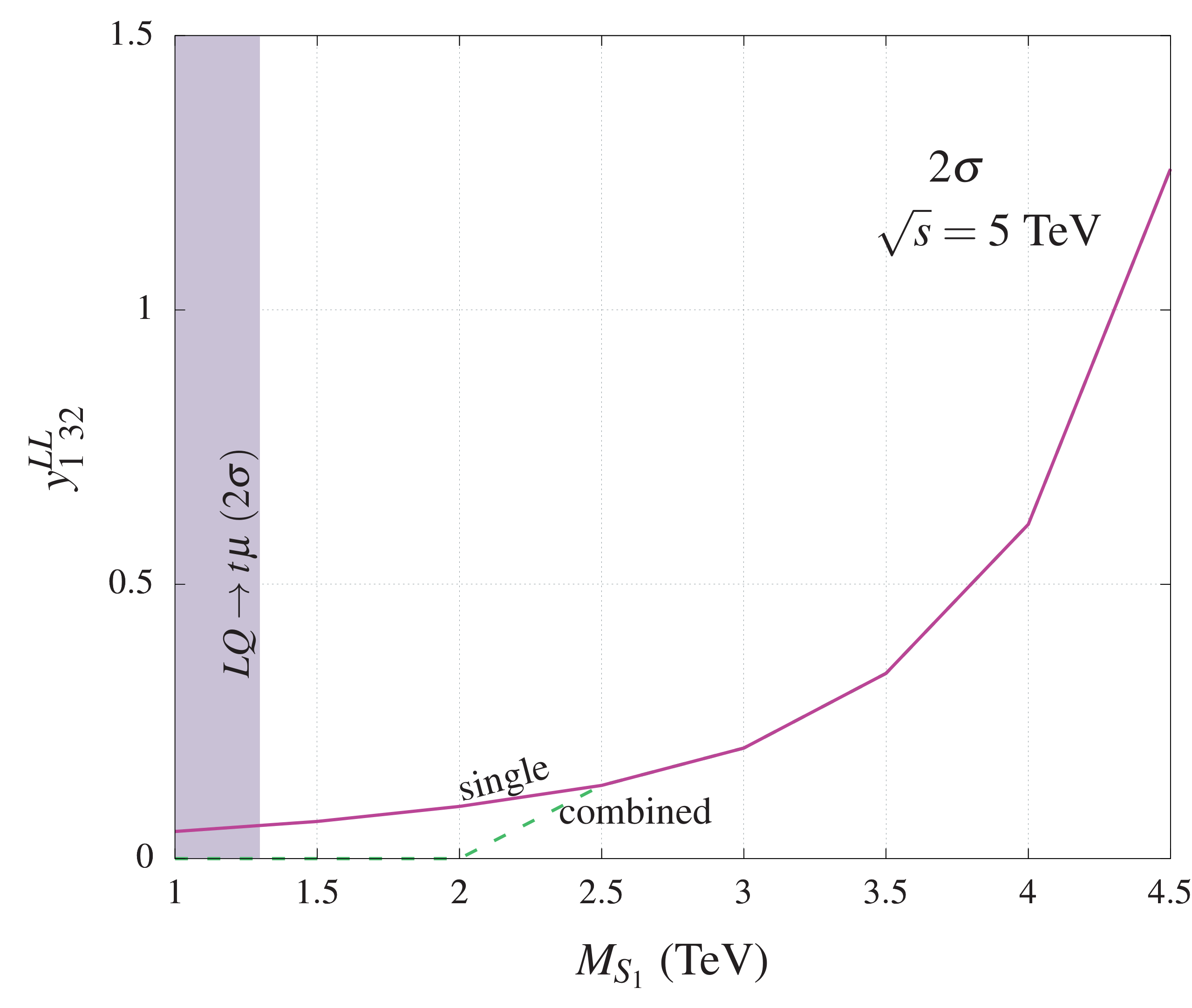}
\label{fig:MS1LQ_lambda_32L_2sig_5TeV}}
\subfloat[\quad\quad\quad(b)]{\includegraphics[width=0.35\linewidth]{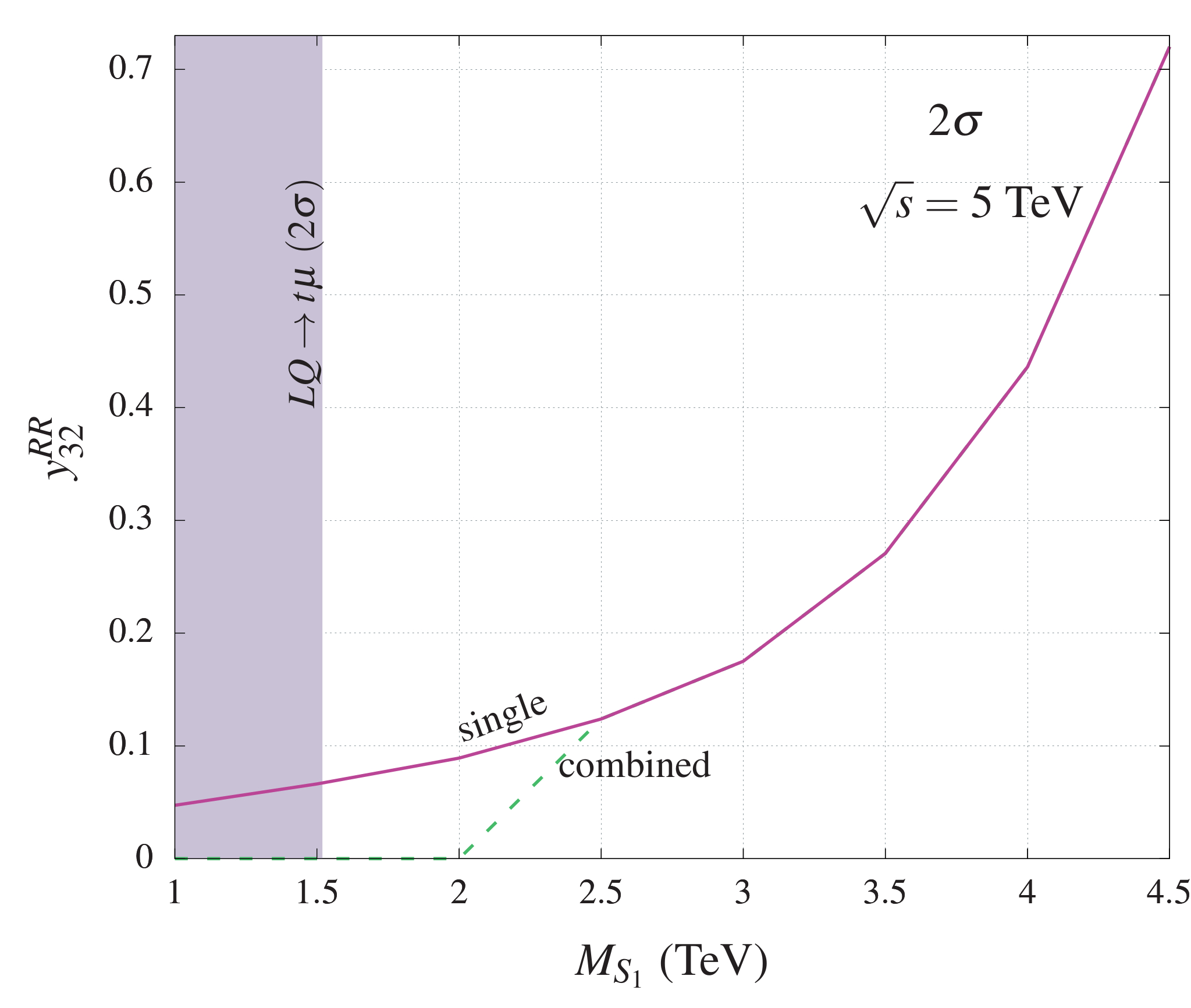}
\label{fig:MS1LQ_lambda_32R_2sig_5TeV}}\\
\subfloat[\quad\quad\quad(c)]{\includegraphics[width=0.35\linewidth]{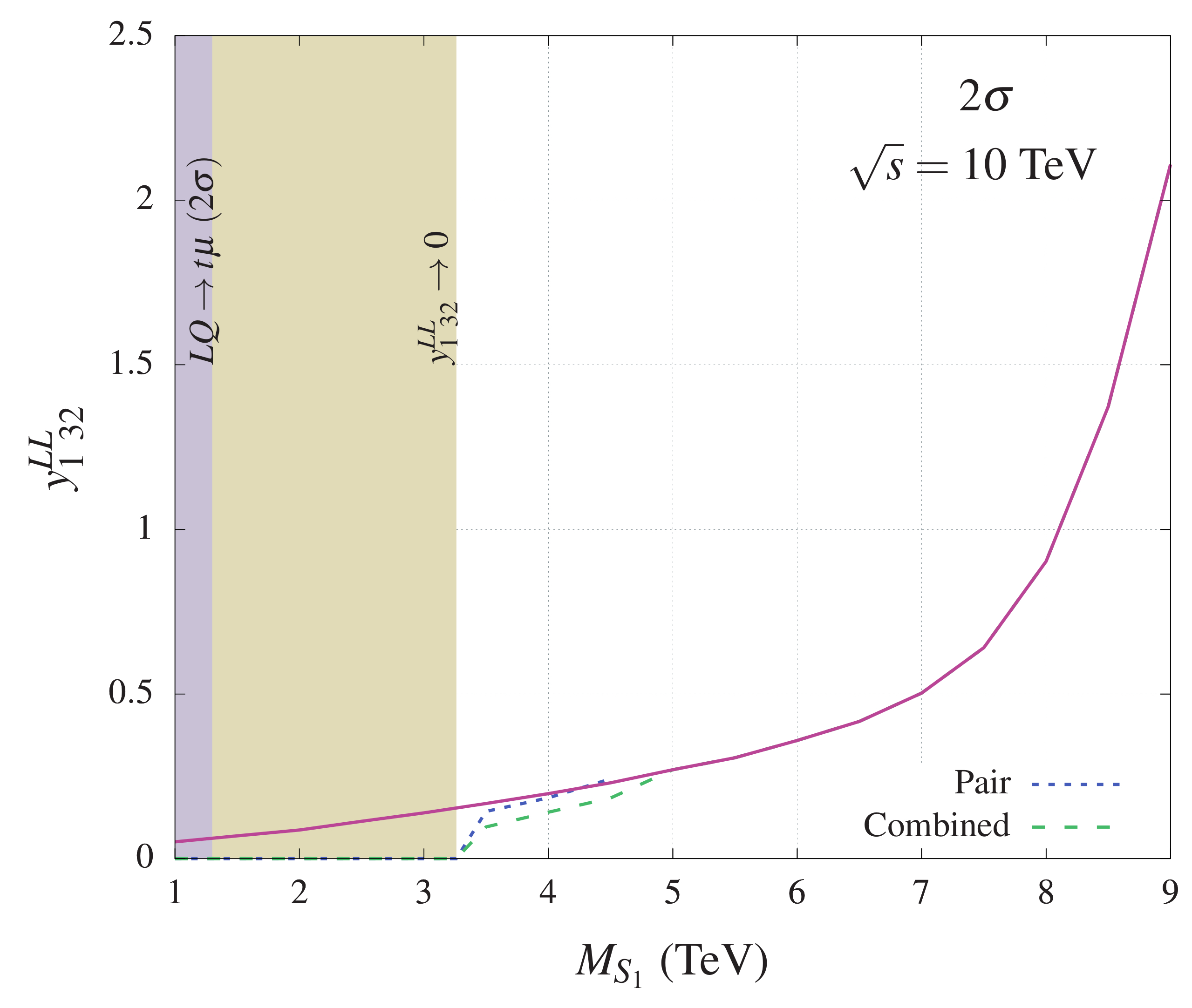}\label{fig:MS1LQ_lambda_32L_2sig_10TeV}}
\subfloat[\quad\quad\quad(d)]{\includegraphics[width=0.35\linewidth]{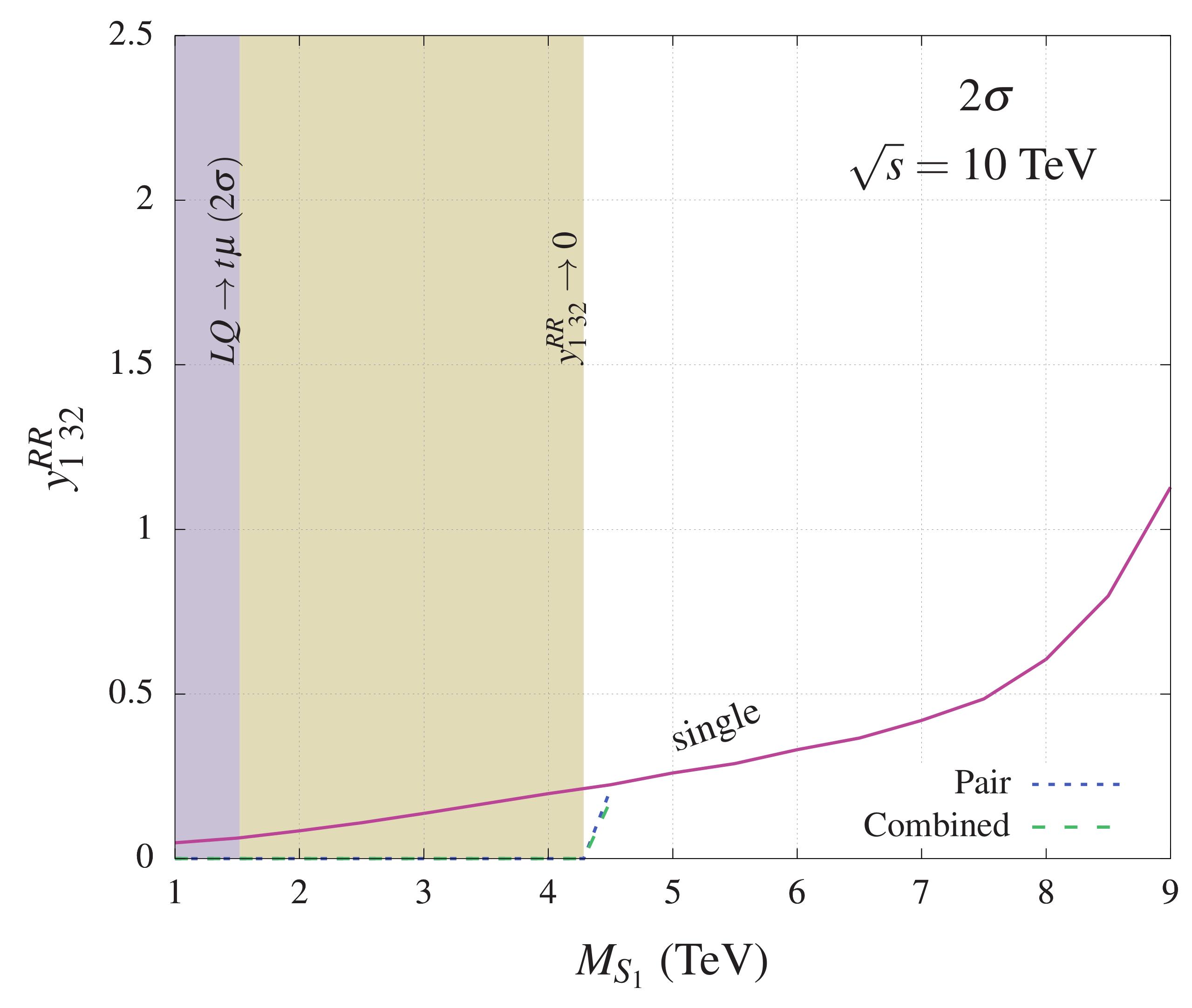}\label{fig:MS1LQ_lambda_32R_2sig_10TeV}}
\caption{The $2\sigma$ exclusion limits as a function of the new yukawa coupling $y_1$ and $M_{S_1}$ (TeV) for benchmark C.O.M energies $5.0$ TeV [(a) and (b)] and $10.0$ TeV [(c) and (d)].}
\label{fig:2sigma}
\end{figure*}


\section{Results and conclusion}
\label{sec:result}
In this work, we have investigated the discovery reach of the sLQ $S_1$ decaying to a top quark and a muon at the proposed muon collider. We systematically combine the contributions from pair and single production of $S_1$ in order to maximise the discovery reach. Our signal topology comprises of atleast one hadronically decaying top quark and exactly two opposite sign muons of which one must have a high $p_T$. We show our results in Figs.~\ref{fig:5sigma} and \ref{fig:2sigma}. In Fig.~\ref{fig:5sigma}, we show the $5\sigma$ discovery reach contours for pair (blue dashed), single (red solid) and combined (green dashed) production modes as function of the mass of $S_1$ and the new coupling $y_1$. These plots shows the least value of $y_1$ required for to observe $S_1$ signal with $5\sigma$ significance for a given mass. The light yellow shaded region denotes the maximum sLQ mass the muon collider can probe with $5\sigma$ significance if the coupling independent pair only mode is considered (See Fig.~\ref{fig:pairproduction2}). The light violet region is excluded by the LHC with $95\%$ confidence limit (C.L) for the LQ searches to top quark and muon\cite{ATLAS:2023prb}. In Figs.~\ref{fig:MS1LQ_LM32L_single_5sig_5TV} [\ref{fig:MS1LQ_LM32R_single_5sig_5TV}], we plot the $5\sigma$ contours for the new coupling $y^{LL}_{1~32}$ [$y^{RR}_{1~32}$] for the $5.0$ TeV benchmark scenario. Similar plots are shown for the $10$ TeV benchmark scenario in Figs.~\ref{fig:MS1LQ_lambda_32L_5sig_10TeV} and \ref{fig:MS1LQ_lambda_32R_5sig_10TeV}.

In Figs.~\ref{fig:MS1LQ_LM32L_single_5sig_5TV} and \ref{fig:MS1LQ_LM32R_single_5sig_5TV}, the pair production contribution ceases beyond the sLQ mass $2.0$ TeV  due to phase space suppression. Post $2.0$ TeV, the combined mode comprises only of the single production process. In Fig.~\ref{fig:MS1LQ_LM32L_single_5sig_5TV}, assuming a small coupling $y_1$, the muon collider can probe a LQ as heavy as $1.6$ TeV (light yellow region). If we assume only the single production mode, the muon collider can probe a sLQ as heavy as $4.5$ TeV with $\mathcal{O}(1)$ new coupling with $5\sigma$ significance. In Fig.~\ref{fig:MS1LQ_lambda_32L_5sig_10TeV} (\ref{fig:MS1LQ_lambda_32R_5sig_10TeV}), assuming a small coupling $y_1$, the discovery reach of the sLQ goes upto $2.1(3.3)$ TeV. The discovery reach is higher for the right handed coupling because the branching ratio of sLQ to top quark and muon is $100\%$ in the right handed scenario ($y^{RR}_{1 32}$). Similarly, if we assume only the single production mode, then the muon collider can probe a LQ as heavy as $9$ TeV with $5\sigma$ significance, while still keeping the new couplings within pertubative limits. In Fig.~\ref{fig:2sigma}, we show similar plots for $2\sigma$ exclusion limits. In the absence of a discovery, these plots show the maximum a sLQ can be excluded with $95\%$ C.L. In Fig.~\ref{fig:MS1LQ_lambda_32R_2sig_10TeV}, we can exclude a sLQ as heavy as $\approx~4.2$ TeV with $95\%$ C.L. if we consider the coupling independent pair production search only. 

These results clearly demonstrate the superiority of a muon collider over the HL-LHC~\cite{Chandak:2019iwj} in probing heavier BSM particles. { The projected sensitivity reach of HL-LHC for similar final state is only about {1.7} TeV at 5$\sigma$ C.L \cite{Chandak:2019iwj}, and hence a muon collider  clearly out-performs the HL-LHC in probing sLQ.} Despite following a cut-based approach for determining the discovery and exclusion, we obtained some promising yet conservative results. In our upcoming work, we plan to utilise sophisticated machine learning (ML) technique to maximise signal background separation and further enhance the discovery potential. Since a ML based approach would enhance the signal efficiency it would allow us to utilise the the nonresonant production mode of the sLQ which otherwise due to its smaller cross section doesn't contribute significantly. This nonresonant production mode is suitable for probing LQs heavier than the C.O.M and for larger couplings.


\acknowledgments 
The authors acknowledge the use of SAMKHYA: High-Performance Computing Facility provided by the Institute of Physics (IOP), Bhubaneswar. M.M acknowledges the IPPP Diva Award research grant.

\bibliography{Leptoquark}{}
\bibliographystyle{JHEPCust}

\end{document}